\documentclass{appolb}
\usepackage{graphicx}
\usepackage{epsfig}
\begin{document}

\title{DAMPING OF GIANT DIPOLE RESONANCE IN HIGHLY EXCITED NUCLEI\thanks{Invited lecture at the Zakopane School on Nuclear Physics "Extremes of the Nuclear Landscape", August 27 - September 2, 2012, Zakopane, Poland.} }

\author{ Nguyen Dinh Dang
\address{Theoretical Nuclear Physics Laboratory, \\ 
Nishina Center for Accelerator-Based Science, \\ 
RIKEN, 2-1 Hirosawa, Wako city, 351-0198 Saitama, Japan \\ and \\ Institute for nuclear science and technique, \\
179 Hoang Quoc Viet, Nghia Do, Hanoi, Vietnam}}
\maketitle

\begin{abstract}
The giant dipole resonance's (GDR) width and shape at finite temperature and angular momentum are described within the phonon damping model (PDM), which predicts an overall increase in the GDR's total width at low and moderate temperature $T$, and its saturation at high $T$. At  $T<$ 1 MeV the GDR width remains nearly constant because of thermal pairing. The PDM description is compared with the experimental systematics obtained from heavy-ion fusion, inelastic scattering of light particles on heavy targets, and $\alpha$ induced fusion reactions, as well as with predictions by other theoretical approaches. The results obtained within the PDM and GDR's experimental data are also employed to predict the viscosity of hot medium and heavy nuclei.
\end{abstract}

  
\section{Introduction}
The giant dipole resonance (GDR) is the best-known fundamental mode of nuclear excitations at high frequencies. The GDR built on the ground state of heavy nuclei has a small width ($\sim$ 4 - 5 MeV) and the integrated cross section up to around 30 MeV that exhausts the Thomas-Reich-Kuhn (TRK) sum rule. The GDR built on highly excited compound (CN) nuclei was observed for the first time in 1981~\cite{Newton}, and at present a wealth of experimental data has been accumulated for the GDR widths at finite temperature $T$ and angular momentum $J$ in various medium and heavy nuclei formed in heavy ion fusions~\cite{Schiller}, deep inelastic scattering of light particles on heavy targets~\cite{Baumann,Heckman}, and $\alpha$ induced fusions~\cite{Kolkata}. The common features of the hot GDR are: (1) Its energy is nearly independent of $T$ and $J$, (2) Its full width at half maximum (FWHM) remains mostly unchanged in the region of $T\leq$ 1 MeV, but increases sharply with $T$ within 1$\leq T\leq$ 2.5 - 3 MeV, and seems to saturate at $T\geq$ 4 MeV. As a function of $J$, a significant increase in the GDR width is seen only at $J\geq$ 25 - 27$\hbar$. In Ref. \cite{Kusnezov} some GDR data were reanalyzed by adding the pre-equlibrium $\gamma$ emission  and it was claimed that the GDR width does not saturate. However, it was realized later that the pre-equilibrium emission is proportional to the asymmetry between projectiles and targets and lowers the CN excitation energy, which alters the conclusion on the role of pre-equlibrium emission. The recent measurements in $^{88}$Mo at $T\geq 3$ MeV and $J>$ 40$\hbar$ did not show any significant effect of pre-equilibrium emission on the GDR width~\cite{Maj}.   The evaporation width owing to the quantal mechanical uncertainty in the energies of the CN states was also proposed to be added into the total GDR width~\cite{Chomaz}. However, the high-energy $\gamma$-ray spectra resulting from the complete {\scriptsize CASCADE} calculations~\cite{Gervais} including the evaporation width turned out to be essentially identical to those obtained by neglecting this width even up to excitation energy higher than 120 MeV for $^{120}$Sn (i.e., at $T >$  3.3 MeV). This indicates that the effect of evaporation width, if any, may become noticeable only at much higher values of $T$ ($\gg$ 3.3 MeV) and $J$ ($\gg$ 30$\hbar$). 
In a classical representation of the GDR as a damped spring mass system, the damping width of the oscillator (the GDR width) should be smaller than its frequency (the GDR energy) otherwise the spring mass system cannot make any oscillation. This means that the GDR width is upper-bounded by its energy. This implies the saturation of the GDR width.

The present lecture summarizes the achievements of the Phonon Damping Model (PDM)~\cite{PDM1a,PDM2,PDMJ} in the description of the the GDR width and shape at finite $T$ and $J$ (Sec. \ref{TJ}). As two applications, the GDR parameters predicted by the PDM and experimentally extracted are used to calculate the shear viscosity of finite hot nuclei, which is also employed to test the recent preliminary data of the GDR width at high $T$ and $J$ in $^{88}$Mo (Sec. \ref{visco}). Conclusions are drawn in the last section.

\section{Damping of GDR in highly excited nuclei}
\label{TJ}
\subsection{GDR width and shape in hot nuclei}
\label{T}
The width of the GDR built of the ground state ($T=$ 0) (the quantal width $\Gamma_{Q}$), consists of the three components: (i) the Landau width $\Gamma^{LD}$, which is essentially the variance $\sqrt{\langle E^2\rangle - \langle E\rangle^2}$ of the $ph$-state distribution, (ii) the spreading width $\Gamma^{\downarrow}$ caused by coupling of $1p1h$ states to more complicated configurations such as $2p2h$ ones, and (iii) the escape width $\Gamma^{\uparrow}$ owing to the direct particle decay into hole states of the residual nucleus because of coupling to continuum. In medium and heavy nuclei, the major contribution to $\Gamma_Q$ is given by $\Gamma^{\downarrow}$, whereas $\Gamma^{LD}$ and $\Gamma^{\uparrow}$ account for a small fraction. The calculations within the microscopic models such as the particle+vibration model~\cite{Bortignon} and the quasiparticle-phonon model~\cite{QPM} have shown that $\Gamma^{\downarrow}$ does not increase width $T$. Therefore the mechanism of the width's increase width $T$ should be sought beyond the one that causes $\Gamma^{\downarrow}$.

The PDM's Hamiltonian consists of the 
independent single-particle (quasiparticle) field, GDR phonon field, and the coupling between 
them [Eq. (1) in Ref. \cite{PDM1a}].   The Woods-Saxon potentials at $T =$ 0 are often 
used to obtain the single-particle energies $\epsilon_k$. The GDR width $\Gamma(T)$ is a sum: $\Gamma(T)=\Gamma_{\rm Q}+\Gamma_{\rm T}$ of
the quantal width, $\Gamma_{\rm Q}$, and thermal width, $\Gamma_{\rm T}$.
In the presence of superfluid pairing, the quantal and thermal widths 
are given as~\cite{PDM2}
\begin{equation}
\Gamma_{\rm Q}=2\gamma_Q(E_{GDR})=2\pi F_{1}^{2}\sum_{ph}[u_{ph}^{(+)}]^{2}(1-n_{p}-n_{h})
\delta[E_{\rm GDR}-E_{p}-E_{h}]~,
\label{GammaQ}
\end{equation}
\begin{equation}
\Gamma_{\rm T}=2\gamma_T(E_{GDR})=2\pi F_{2}^{2}\sum_{s>s'}[v_{ss'}^{(-)}]^{2}(n_{s'}-n_{s})
\delta[E_{\rm GDR}-E_{s}+E_{s'}]~, 
\label{GammaT}
\end{equation}
where $u_{ph}^{(+)} = u_pv_h+u_hv_p$, $v_{ss'}^{(-)}=u_su_{s'}-v_sv_{s'}$ ($ss' = pp', hh'$) with
$u_k$ and $v_k$ being the coefficients of Bogolyubov's transformation from particle operators to 
quasiparticle ones, $E_k\equiv\sqrt{(\epsilon_k-\lambda)^2+\Delta^2}$, with superfluid pairing gap $\Delta$,
are quasiparticle energies, $n_k$ are quasiparticle occupations numbers, which, for medium and heavy nuclei, can be well approximated with the Fermi-Dirac distribution for
independent quasiparticles, $n_k = [\exp(E_k/T)+1]^{-1}$. The parameter $F_1$ is chosen so that $\Gamma_Q$ at $T=$ 0 is equal to GDR's width at $T=$ 0, whereas the parameter $F_2$ is chosen so that, with varying $T$,  the GDR energy $E_{GDR}$ does not change significantly. The latter is found as the solution of the equation $E_{GDR} - \omega_{q}-P_q(E_{GDR})=0$, where $\omega_q$ is the energy of the GDR phonon before the coupling between the phonon and single-particle mean fields is switched on, and $P_q(\omega)$ is the polarization operator owing to this coupling, whose explicit expression in given in Refs. \cite{PDM2}. The GDR strength function is calculated as 
\begin{equation}
S_{q}(\omega) = \frac{1}{\pi}\frac{\gamma_Q(\omega) + \gamma_T(\omega)}{(\omega-E_{GDR})^2+[\gamma_Q(\omega) + \gamma_T(\omega)]^{2}}~.\label{S}
\end{equation}
In numerical calculations the representation $\delta(x) =\lim_{\varepsilon\rightarrow 0}\varepsilon/[\pi(x^{2}+\varepsilon^2)]$ is used for the $\delta$-functions in Eqs. (\ref{GammaQ}) and (\ref{GammaT}) with $\epsilon=$ 0.5 MeV.

\begin{figure}
\center{
\includegraphics[width=10.3cm]{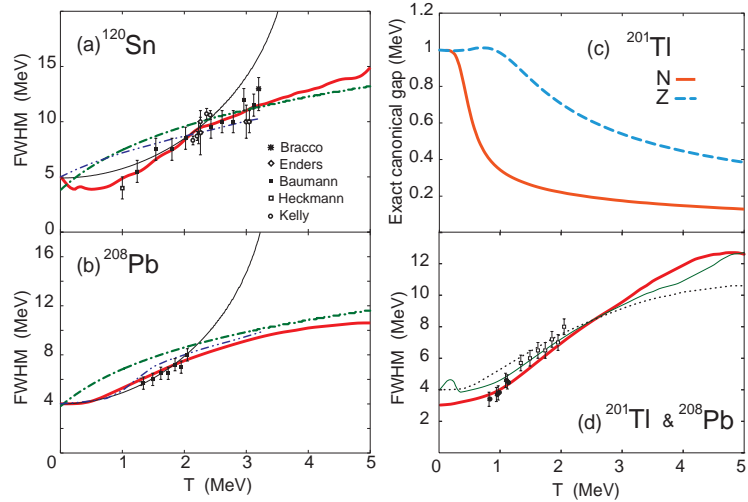}}
\caption{GDR widths for $^{120}$Sn (a) and $^{208}$Pb (b) predicted by the PDM (thick solid), pTSFM (dot-dashed), AM (double dot-dashed), and FLDM (thin solid) as functions of $T$ in comparison with experimental data in tin and lead regions. (c): Exact canonical neutron (N) and proton (Z) pairing gaps for $^{201}$Tl as functions of $T$. (d): GDR width for $^{201}$Tl obtained within the PDM as a function of $T$ (thick solid) including the exact canonical gaps in (c) in comparison with the experimental data for $^{201}$Tl (black circles) and $^{208}$Pb (open boxes). The thin solid line is the PDM result without the effect of thermal pairing. The dotted line is the PDM result for $^{208}$Pb [the same as the thick solid line in (b)]. 
\label{widthSnPb}}
\end{figure}
The GDR widths predicted by the PDM, the two versions of thermal shape fluctuation model (TSFM), namely the phenomenological TSFM (pTSFM)~\cite{pTSFM} and the adiabatic model (AM)~\cite{AM}, and the Fermi liquid drop model (FLDM)~\cite{FLDM} for $^{120}$Sn and $^{208}$Pb are shown in Figs. \ref{widthSnPb} (a) and \ref{widthSnPb} (b) in comparison with the experimental systematics. The PDM results for $^{120}$Sn include the effect of non-vanishing thermal pairing gap because of thermal fluctuations owing to finiteness of nuclei. The figure clear shows that among the models under consideration, the PDM is the only one that is able to describe well the experimental data in the entire temperature region including $T\leq$ 1 MeV, where the other model fail. It is also able to reproduce the very recent data for the GDR width in $^{201}$Tl at 0.8$\leq T<$ 1.2 MeV [Fig. \ref{widthSnPb} (d)] after including the exact canonical 
gaps for neutrons and protons shown in Fig. \ref{widthSnPb} (c)~\cite{Tl201}.

\begin{figure}
\center{
\includegraphics[width=9.3cm]{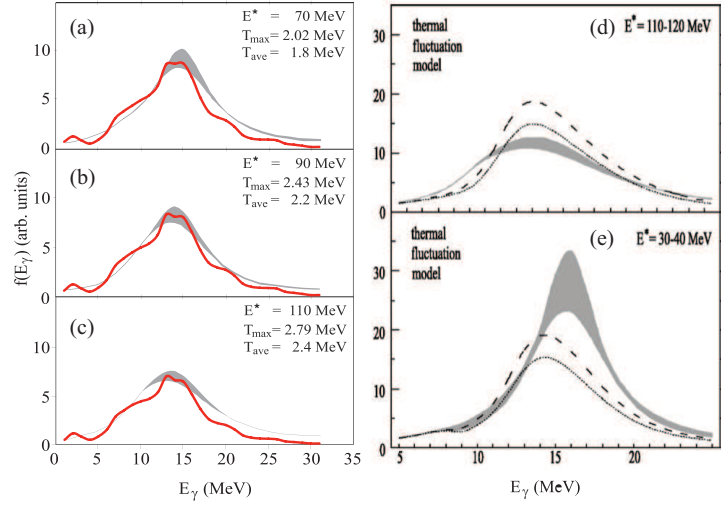}}
\caption{Experimental (shaded areas) and theoretical shapes for GDR in $^{120}$Sn generated by the {\scriptsize CASCADE} code at various excitation energies. (a) -- (c): predictions generated by using the PDM strength functions. (d) and (e)~\cite{Gervais}: dotted and dashed lines are generated by using the TSFM strength functions, exhausting 80$\%$ and 100$\%$ TRK sum rule, respectively.\label{SE}}
\end{figure}
For an adequate description of not only the width but also the entire GDR shape, the PDM strength functions were incorporated into all the decay steps of the full statistical calculations and the generated results are compared with those obtained from the measured $\gamma$-ray spectra in Fig. \ref{SE}, which shows that the PDM describes fairly well the GDR shape [Fig. \ref{SE} (a)], whereas the TSFM fails in doing so [Fig. \ref{SE} (b)].
\subsection{GDR width and shape in hot and rotating nuclei}
\label{J}
To describe the non-collective rotation of a spherical nucleus, the $z$-projection $M$ of the total angular momentum $J$ is added into the PDM Hamiltonian as  $- \gamma\hat{M}$, where $\gamma$ is the rotation frequency~\cite{PDMJ}. The latter and the chemical potential are defined, in the absence of pairing,  from the equation $M = \sum_k m_k(f_{k}^{+}-f_{k}^{-})~,$ and  $N =\sum_k(f_{k}^{+}+f_{k}^{-})~$, where $N$ is the particle number and $f_k^{\pm}$ are the single-particle occupation numbers, $f_{k}^{\pm} =1/[\exp(\beta E_k^{\mp})+1]$, and $E_k^{\mp} = \epsilon_k-\lambda\mp\gamma m_k~$. With the smoothing of $\delta$-functions by using the Breit-Wigner distribution mentioned in Sec. \ref{T}, the final form of phonon damping $\gamma_q(\omega)$ becomes
\begin{equation}
\gamma_q(\omega) = \varepsilon\sum_{kk'}[{\cal F}_{kk'}^{(q)}]^{2}
\bigg[\frac{f_{k'}^{+}-f_{k}^{+}}
{(\omega-E_k^{-} + E_{k'}^{-})^2+\varepsilon^2}+\frac{f_{k'}^{-}-f_{k}^{-}}{(\omega-E_k^{+} + E_{k'}^{+})^{2}+\varepsilon^{2}}\bigg]~,
\label{gamma1}
\end{equation}
where $(k,k') = ph, pp', hh'$. The GDR strength function is calculated by using the same Eq. (\ref{S}) where 
$\gamma_Q(\omega) +\gamma_T(\omega)$ is replaced with $\gamma_q(\omega)$. The explicit expression for the polarization operator $P_q(\omega)$ is given in Eq. (13) of Ref. \cite{PDMJ}.

\begin{figure}
     \center{\includegraphics[width=9cm]{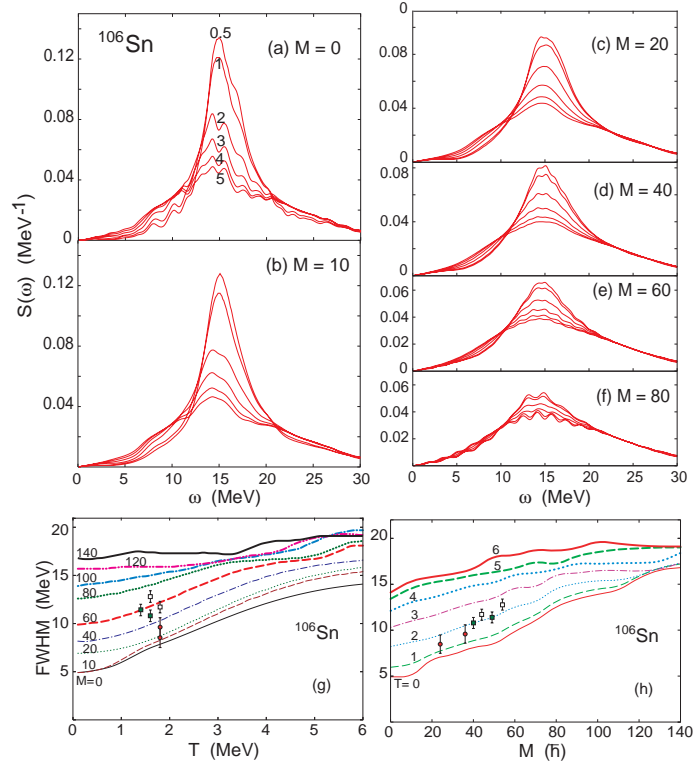}
     \caption{(a) -- (f): GDR strength functions  for
     $^{106}$Sn at $T=$ 0.5, 1, 2, 3, 4, and 5 MeV as shown at the curves and $M=$ 0, 10, 20, 40, 60, and 80$\hbar$ as shown in the panels.
     (g): FWHM of GDR for $^{106}$Sn as a function of $T$
     at several values of $M$ (in $\hbar$) shown at the curves, (h): FWHM of GDR for $^{106}$Sn as a function of $M$ at several values of $T$ (in MeV) shown at the curves. The experimental data for GDR in $^{106}$Sn (solid circles), $^{109,110}$Sn (solid and open boxes) are adapted from Refs. \cite{Mattiuzzi,Bra}.}    
      \label{streJSn}}
\end{figure}
Shown in  Fig. \ref{streJSn}  are the GDR strength functions $S(\omega)$ and the widths in
$^{106}$Sn at various $T$ and $M$.  The GDR shape becomes smoother  as $T$ and $M$ increase, and the smoothing caused by the angular momentum is stronger than that caused by thermal effects (Figs. \ref{streJSn} (a) -- \ref{streJSn} (f)). The GDR width increases with both $T$ and $M$ and stronger at low $T$ and $M$. This increase in the width approaches a saturation at moderate and high $T$ and/or $M$. As a function of $T$, the saturation begins at $T>$ 4 MeV in $^{106}$Sn [Fig. \ref{streJSn} (g)], whereas as a function of $M$ it takes place in $^{106}$Sn already at $T\geq$ 3 MeV [Fig. \ref{streJSn} (h)].  Experimental data for $^{106}$Sn~\cite{Mattiuzzi} and $^{109,110}$Sn~\cite{Bra} are also shown in Figs. \ref{streJSn} (g) and \ref{streJSn} (h), which are in fair agreement with theory.
\section{Shear viscosity of hot nuclei}
\label{visco}
In the verification of the condition for applying hydrodynamics to 
nuclear system, the quantum mechanical 
uncertainty principle requires a finite viscosity for any thermal 
fluid. Kovtun, Son and Starinets (KSS)~\cite{KSS} conjectured
that the ratio $\eta/s$ of shear viscosity $\eta$ to the entropy 
volume density $s$ is bounded below for all fluids, namely the value
${\eta}/{s}= {\hbar}/(4\pi k_{B})$  
is the universal lower bound (KSS bound or unit).
From the viewpoint of collective theories, one of the fundamental 
explanations for the giant resonance damping is the friction term (or viscosity) 
of the neutron and proton fluids.  By using the Green-Kubo's relation, it has been shown in Ref. \cite{visco} that the shear viscosity $\eta(T)$ 
at finite $T$ is expressed in terms of the GDR's parameters at zero and finite $T$ as
\begin{equation}
\eta(T)=\eta(0)\frac{\Gamma(T)}{\Gamma(0)}
\frac{E_{GDR}(0)^{2}+[\Gamma(0)/2]^{2}}{E_{GDR}(T)^{2}+[\Gamma(T)/2]^{2}}~.
\label{eta1}
\end{equation}
\begin{figure}
     \center{\includegraphics[width=9.0cm]{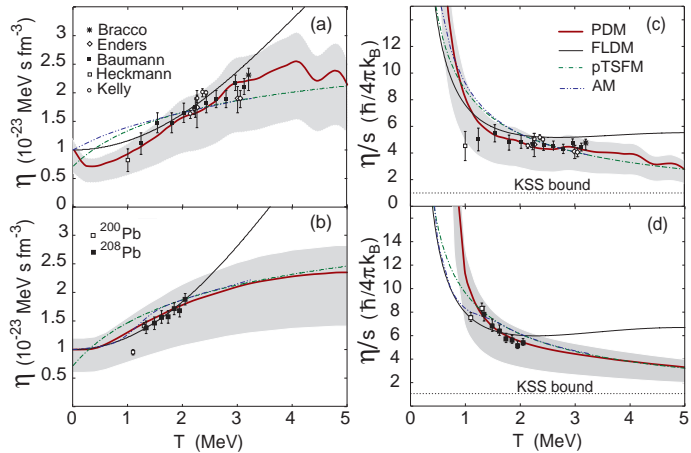}
     \caption{Shear viscosity $\eta(T)$ [(a) and (b)] and ratio 
     $\eta/s$ [(c) and (d)] as functions of $T$ for nuclei in tin [(a) and (c)], and lead [(b) and (d)] regions. The  
     gray areas are the PDM predictions by using $0.6u\leq\eta(0)\leq 
     1.2u$ with $u=$ 10$^{-23}$ Mev s fm$^{-3}$.}
     \label{eta&ratio}}
\end{figure}
The predictions for the shear viscosity $\eta$ and the ratio $\eta/s$ by the PDM,  pTSFM, AM, and 
 FLDM for $^{120}$Sn and $^{208}$Pb are plotted as functions of $T$ 
in Fig. \ref{eta&ratio} in comparison with the empirical results. 
The latter are extracted from the 
experimental systematics for GDR in tin and lead regions~
\cite{Schiller} making use of Eq. (\ref{eta1}).  It is seen in Fig. \ref{eta&ratio} that the predictions by the PDM have the best overall 
agreement with the empirical results. It produces an increase 
of $\eta(T)$ with $T$ up to 3 - 3.5 MeV and a saturation of $\eta(T)$ within (2 - 3)$u$ 
at higher $T$ [with 
$\eta(0)=$ 1$u$, $u=$ 10$^{-23}$ Mev s fm$^{-3}$]. The ratio $\eta/s$ decreases sharply with 
increasing $T$ up to $T\sim$ 1.5 MeV, starting from which the decrease 
gradually slows down to reach (2 - 3) KSS units 
at $T=$ 5 MeV. The FLDM has a similar trend as that of the PDM up to 
$T\sim$ 2 - 3 MeV, but at higher $T$ ($T>$ 3 MeV for $^{120}$Sn or 2 MeV for 
$^{208}$Pb) it produces an increase of both $\eta$ and $\eta/s$ with $T$. 
At $T=$ 5 MeV the FLDM model predicts the ratio $\eta/s$ within (3.7 - 6.5) KSS units, which are 
roughly 1.5 -- 2 times larger than the PDM predictions. The AM and pTSFM show a similar trend for $\eta$ and $\eta/s$. 
However, in order to obtain such similarity, $\eta(0)$ in the pTSFM 
calculations has to be reduced to 0.72$u$ instead of 1$u$. They all 
overestimate $\eta$ at $T<$ 1.5 MeV. 

A model-independent estimation for the high-$T$ limit of the ratio $\eta/s$ can also be inferred 
directly from Eqs. (\ref{eta1}). Assuming that,
at the highest $T_{max}\simeq$ 5 - 6 MeV where the GDR can still exist, the 
GDR width $\Gamma(T)$ cannot exceed $\Gamma_{max}\simeq 
3\Gamma(0)\simeq 0.9E_{GDR}(0)$~\cite{Auerbach1}, and $E_{GDR}(T)\simeq E_{GDR}(0)$, one obtains from 
Eq. (\ref{eta1}) $\eta_{max} \simeq 2.551\times\eta(0)$. By noticing 
that, $S_{F}\to 2\Omega\ln{2}$ at $T\to\infty$ 
because $n_{j}\to$ 1/2, where $\Omega=\sum_{j}(j+1/2)$ for 
the spherical single-particle basis or sum of all doubly-degenerate levels for the 
deformed basis and that the particle-number conservation requires $A = 
\Omega$ since all single-particle occupation numbers are equal to 
1/2, one obtains the high-$T$ limit of entropy density $s_{max} = 2\rho\ln{2}\simeq 0.222~(k_{B})$.
Dividing $\eta_{max}$ by $s_{max}$ yields the high-$T$ limit (or lowest bound) for $\eta/s$ 
in finite nuclei, that is $({\eta}/{s})_{min}\simeq 2.2^{+0.4}_{-0.9}$ KSS units, where the empirical values for $\eta(0) = 
1.0^{+0.2}_{-0.4}~u$ are used~\cite{Auerbach1,fission}. Based on these results, one can conclude that
the value of $\eta/s$ for medium and heavy nuclei at $T=$ 5 MeV is in 
between (1.3 - 4.0) KSS units, which is about (3 - 5) times smaller 
(and of much less uncertainty) that the value between (4 - 19) KSS units predicted by 
the FLDM for heavy nuclei~\cite{Auerbach}, where the same lower value $\eta(0)=$0.6$u$ was used.

\begin{figure}
\center{
     \includegraphics[width=11 cm]{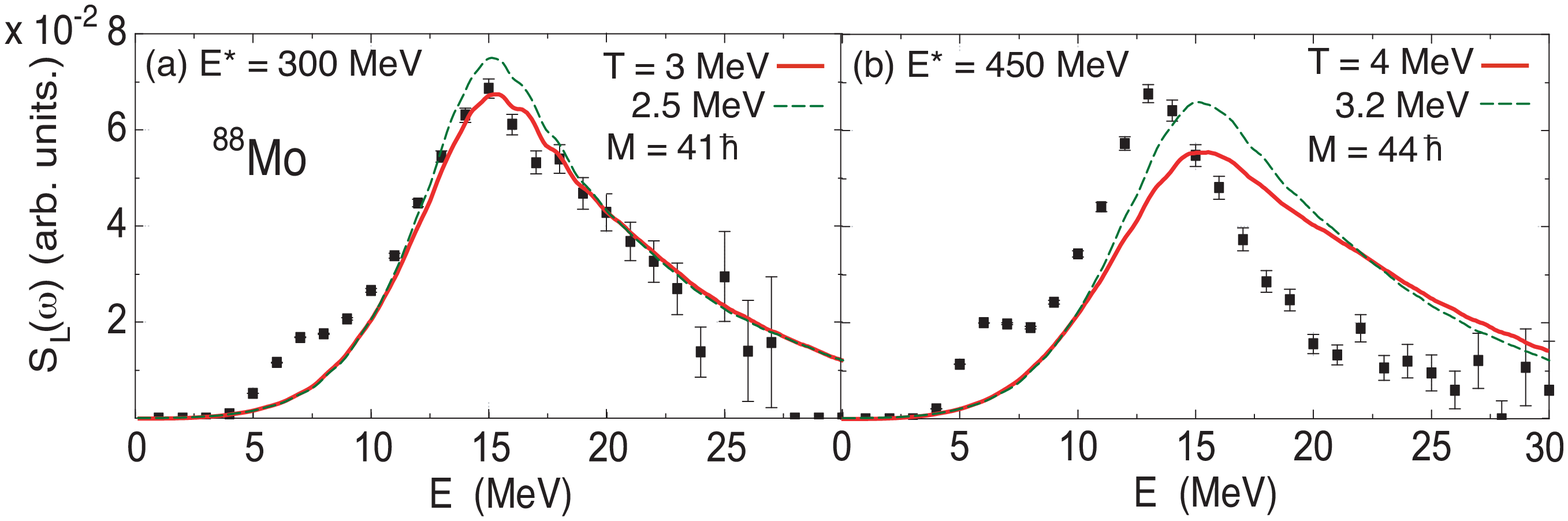}
     \caption{GDR strength function $S_L(\omega)$ for $^{88}$Mo at $M=$ 41 $\hbar$ (a) and 
    $M=$ 44 $\hbar$ (b) predicted by the PDM in comparison with the preliminary data from Ref. \cite{Maj}.}     
      \label{Stest}}
\end{figure}

Finally, by using the temperature dependence of $\eta/s$ and the KSS lower bound conjecture, it is possible to examine the recent preliminary data for the GDR width in $^{88}$Mo in Ref. \cite{Maj}. Shown in Fig. \ref{Stest} is the strength function 
$S_L(\omega) = {\omega}[S(\omega, E_{GDR})-S(\omega, -E_{GDR})]/E_{GDR}$, where $S(\omega, \pm E_{GDR})$ are the PDM strength functions (\ref{S}) at finite $T$ and $J$ for the GDR located at $\pm E_{GDR}$. The PDM predictions are shown at the initial temperature $T_{max}$ of the compound nucleus ($T_{max} =$ 3 and 4 MeV in Figs. \ref{Stest} (a) and \ref{Stest} (b), respectively), and also at $T=$ 2.5  MeV (Figs. \ref{Stest} (a)) and 3.2 MeV (Figs. \ref{Stest} (b)), i.e. within the error bars of the average temperature $\langle T\rangle$ obtained by averaging over all the GDR decay steps ($\langle T\rangle =$ 2$\pm$0.6 and 2.6$\pm$0.8 MeV for $E^{*}=$ 300 and 450 MeV, respectively~\cite{Maj1}). While the PDM strength functions and experimental line shapes of the GDR agree fairly well at $M=$ 41 $\hbar$ with the FWHM $\Gamma$ predicted by the PDM between 9.6 MeV (T = 2.5 MeV) and 11 MeV (T = 3 MeV), they strongly mismatch at $M=$ 44 $\hbar$, where the experimental GDR peak becomes noticeably narrower with a width $\Gamma_{ex}\simeq$ 7.5 MeV. By using this value $\Gamma_{ex}$ and $\eta(0)=$ 0.6 $u$, one ends up with the value of $\eta/s=$ 0.85 KSS units. Including the error bars in $\Gamma_{ex}$ leads to $\Gamma_{ex}^{<}\simeq$ 6 MeV and $\Gamma_{ex}^{>}\simeq$ 8.5 MeV, which give the values of $\eta/s$ equal to 0.69 and 0.94 KSS, respectively. All these values are smaller than the KSS lower bound conjecture. This may indicate that either (i) the data analysis in extracting the experimental GDR strength function for $^{88}$Mo at $E^{*} = 450$ MeV (Fig. \ref{Stest} (b)) is inaccurate, or (ii) a violation of the KSS conjecture has been experimentally confirmed for the first time ever. The reanalysis of the data is now underway to clarify which one from these two conclusions holds~\cite{Maj1}. 

\section{Conclusions}
The PDM generates the damping of GDR through its couplings to $ph$ configurations, causing the quantal width, as well as to $pp$ and/or $hh$ configurations, causing the thermal width. This leads to an overall increase in the GDR width at low and moderate $T$, and its saturation at high $T$. At very low T $<$ 1 MeV the GDR width remains nearly constant because of thermal pairing. The GDR width also increases with angular momentum $M$ and saturates at high $M$, but this saturation goes beyond the value of maximal angular momentum that the nucleus can sustain without violating the KSS conjecture, that is 46 and 55$\hbar$ for $^{88}$Mo and $^{106}$Sn, respectively, if the value $\eta(0) =$ 0.6$\times 10^{-23}$ Mev s fm$^{-3}$ for the shear viscosity at $T=$ 0 is used.   The PDM predictions agree well with the experimental systematics for the GDR width and shape in various medium and heavy nuclei. The PDM also predicts the shear viscosity to the entropy-density ratio $\eta/s$ between (1.3 - 4.0) KSS units for medium and heavy nuclei at $T=$ 5 MeV, almost the same at that of the quark-gluon-plasma like matter at $T>$ 170 MeV discovered at RHIC and LHC. The PDM and the KSS conjecture are also used to show that the recent preliminary experimental data for GDR in $^{88}$Mo~\cite{Maj} need to be reanalyzed. 
    
\end{document}